# Magnetic graphs for cavity quantum electrodynamics

Sunkyu Yu[1†], Xianji Piao[2§], and Namkyoo Park[3*]

[1]Intelligent Wave Systems Laboratory, Department of Electrical and Computer Engineering, Seoul National University, Seoul 08826, Korea

[2]Wave Engineering Laboratory, School of Electrical and Computer Engineering, University of Seoul, Seoul 02504, Korea

[3]Photonic Systems Laboratory, Department of Electrical and Computer Engineering, Seoul National University, Seoul 08826, Korea Seoul 08826, Korea

E-mail address for correspondence: sunkyu.yu@snu.ac.kr; piao@uos.ac.kr; nkpark@snu.ac.kr

**Abstract**

Strengthening light–matter coupling has become a central challenge in cavity quantum electrodynamics (QED), enabling ultrafast gate operations, qubit protection, and deterministic nonlinear optics. As the coupling increases, even the simplest configuration—a two-level atom interacting with a quantized field—requires careful treatment, as exemplified by the gauge-invariant quantum Rabi model (QRM). Here we propose a magnetic graph model for single-atom cavity QED, which enables the interpretation of quantum dynamics across the ultrastrong coupling regime through graph connectivity. We demonstrate that the generalized QRM maps onto a complex bipartite graph of identical sites under Floquet boundary conditions. This framework captures the crossover from weak to deep-strong coupling via a single metric—the cost of disconnecting a nonmagnetic subgraph. We examine the mechanism underlying this connectivity transition, establishing phase frustration induced by subgraph topology as the primary driver. Scalable to many-body systems, this approach bridges graph theory and cavity QED, revealing highly complex-graph dynamics even in the simplest setting.

## Introduction

Single-atom light–matter interaction is a cornerstone of cavity quantum electrodynamics (QED). However, its theoretical framework—the quantum Rabi model (QRM)—remains under active development[1-7], particularly regarding its validity in the ultrastrong coupling (USC; $0.1 \lesssim \eta \lesssim 1$) and deep-strong coupling (DSC; $\eta \gtrsim 1$) regimes[8,9], where $\eta = g/\omega_0$ is the dimensionless coupling strength, with $g$ the coupling and $\omega_0$ the cavity frequency. The intricacies of handling the QRM are intertwined with the unprecedented dynamics observed in the USC and DSC regimes: the Schrödinger's cat ground states[10], nonlinear optics with virtual photons[11], photon excitation using mechanical engineering[3], and dressed-picture dissipation[12].

The traditional $\eta$-based classification of the USC and DSC regimes is tied to the validity of specific approximations[13]. In the dipole-gauge formulation, the USC regime is marked by the substantial influence of counter-rotating terms, which can be neglected under the rotating-wave approximation (RWA) for weak and strong coupling regimes. By contrast, the DSC dynamics are well captured by an adiabatic displaced-oscillator picture, while the RWA collapses completely.

Recent insights grounded in the gauge principle have highlighted the need for an alternative framework to understand the USC and DSC regimes[1,5-7,14-16]. The gauge-invariant, Coulomb-gauge formulation of the QRM[1]—obtained by sequentially applying truncation and minimal coupling replacement—leads to the appearance of field operators of all orders, obscuring the conventional USC-DSC classification derived from earlier Coulomb- or dipole-gauge formulations. These higher-order interactions, which mediate long-range hopping between states, invalidate a simple description of the RWA developed under examining at most quadratic field operators. Simultaneously, the presence of long-range hopping inspires a graph-based perspective by analogy with shortcuts in complex networks[17].

Here we show that a semi-infinite, weighted magnetic graph having topological features is hidden within the simplest quantum system. Starting from the generalized QRM, we develop its magnetic graph interpretation, revealing that the USC and DSC regimes can be classified by a graph connectivity metric—the cost of isolating a large magnetically trivial subgraph. We analyse how topological features and connectivity bottlenecks contribute to this cost, demonstrating the critical impact of phase frustration in quantum state localization. Our proposal provides metrics that not only classify cavity QED via quantum state connectivity but also probe topological features in hybrid quantum systems.

## Results

### Gauge-invariant QRM

To develop a graph model for cavity QED, we start with a two-level atom coupled to a quantized cavity field (Fig. 1a). We employ the gauge-invariant QRM, which provides a consistent description of time-dependent systems, particularly in the USC and DSC regimes[1]. We extend the Hamiltonian $H_C$ in the earlier work[1] to encompass a vector gauge field $\mathbf{A} = \mathbf{A}_0(a^\dagger + a)$ (Fig. 1a) and generalized dipole moments that accommodate atoms with broken parity or time-reversal symmetry (Fig. 1b).

Consider an atom described by a bare Hamiltonian $H_0 = h_0 I + \mathbf{h}\cdot\boldsymbol{\sigma}$ in the two-level subspace spanned by an orthonormal basis $\{|u_0\rangle, |u_1\rangle\}$, where $h_0$ and $\mathbf{h}$ are a real-valued scalar and vector, respectively, $\boldsymbol{\sigma} = \mathbf{i}\sigma_x + \mathbf{j}\sigma_y + \mathbf{k}\sigma_z$ denotes the Pauli vector, and $I$ is the identity matrix. The basis dependence is also encoded in the permanent dipole moments $\mathbf{d}_\pm = (\mathbf{d}_{11} \pm \mathbf{d}_{00})/2$, and the transition dipole moment $\mathbf{d} = \mathbf{d}_\mathrm{r} + i\mathbf{d}_\mathrm{i}$ with $\mathbf{d}_\mathrm{r} = \mathrm{Re}[\mathbf{d}_{10}]$ and $\mathbf{d}_\mathrm{i} = \mathrm{Im}[\mathbf{d}_{10}]$, where $\mathbf{d}_{rs} = q\langle u_r|\mathbf{x}|u_s\rangle$ (End Matter

A). By applying the minimal coupling replacement[1], the QRM Hamiltonian becomes (End Matter B):

$$H_{\mathrm{C}} = \hbar\omega_0 a^\dagger a + h_0 I + \left[\mathbf{h}_{\parallel} + \mathbf{h}_{+} D(+2i\eta) + \mathbf{h}_{-} D(-2i\eta)\right]\cdot\boldsymbol{\sigma}, \tag{1}$$

where $\eta = |\mathbf{g}|/\hbar$ is the dimensionless coupling strength in the Coulomb gauge obtained with $\mathbf{g} = (\mathbf{A}_0\cdot\mathbf{d}_{\mathrm{r}})\mathbf{i} - (\mathbf{A}_0\cdot\mathbf{d}_{\mathrm{i}})\mathbf{j} + (\mathbf{A}_0\cdot\mathbf{d}_{-})\mathbf{k}$, $D(\alpha)$ denotes the $\alpha$-displacement operator, $\mathbf{h}_{\parallel}$ and $\mathbf{h}_{\perp}$ are the components of $\mathbf{h}$ parallel and perpendicular to $\mathbf{g}$, respectively, and $\mathbf{h}_{+}$ and $\mathbf{h}_{-}$ are defined as

$$\mathbf{h}_{\pm} = \frac{1}{2}\left(\mathbf{h}_{\perp} \mp i\mathbf{h}\times\frac{\mathbf{g}}{|\mathbf{g}|}\right). \tag{2}$$

Notably, $\hat{\mathbf{h}}_{\pm} = \mathbf{h}_{\pm}/|\mathbf{h}_{\perp}|$ and $\hat{\mathbf{h}}_{\parallel} = \mathbf{h}_{\parallel}/|\mathbf{h}_{\parallel}|$ constitute an orthogonal basis, leading to the spin basis states $|\varphi_{\pm}\rangle$ that satisfy $(\hat{\mathbf{h}}_{\parallel}\cdot\boldsymbol{\sigma})|\varphi_{\pm}\rangle = \pm|\varphi_{\pm}\rangle$, $(\hat{\mathbf{h}}_{\pm}\cdot\boldsymbol{\sigma})|\varphi_{\pm}\rangle = O$, and $(\hat{\mathbf{h}}_{\mp}\cdot\boldsymbol{\sigma})|\varphi_{\pm}\rangle = \pm|\varphi_{\mp}\rangle$.

By employing the eigenbasis $\{|u_0\rangle, |u_1\rangle\}$ (Fig. 1b), cavity QED with $\mathbf{h} = h_z\mathbf{k}$ is governed solely by $\mathbf{g}$, which determines $\eta$—the conventional parameter used to classify the USC and DSC regimes[1,13]. While a nonzero $\mathbf{d}_{-}$ indicates broken parity symmetry, broken time-reversal symmetry prevents cancellation of $\mathbf{d}_{\mathrm{i}}$. The vector $\mathbf{A}_0(t)$ accounts for dynamical cavity modulations (Fig. 1a). Consequently, Eq. (1) captures a broad class of realistic scenarios.

**Floquet Rabi graph**

Inspired by $D(\alpha)$, which connects distant number states, we develop a graph model for the generalized QRM, by employing the Fock-state lattice representation[18-21]. For the Schrödinger equation $i\hbar|\psi\rangle = H_{\mathrm{C}}|\psi\rangle$, we use the ansatz $|\psi\rangle = \sum_n(c_n^{+}|\varphi_{+}\rangle + c_n^{-}|\varphi_{-}\rangle)|n\rangle$, where $|n\rangle$ denotes the number state. Substituting this ansatz into the Schrödinger equation yields (End Matter C):

$$i\hbar\partial_t c_n^+ = \left(n\hbar\omega_0 + h_0 + |\mathbf{h}_\parallel|\right)c_n^+ + |\mathbf{h}_\perp|\sum_{m=0}^{\infty} l_{nm}^+ c_m^-,$$
$$i\hbar\partial_t c_n^- = \left(n\hbar\omega_0 + h_0 - |\mathbf{h}_\parallel|\right)c_n^+ + |\mathbf{h}_\perp|\sum_{m=0}^{\infty} l_{nm}^- c_m^+, \tag{3}$$

where $l_{nm}^\pm = \langle n|D(\pm 2i\eta)|m\rangle$ determines the hopping between the sites $|\varphi_\pm\rangle|n\rangle$ and $|\varphi_\mp\rangle|m\rangle$ (Fig. 1c), evaluated as

$$l_{nm}^\pm = e^{-2\eta^2}\sqrt{\frac{\min\{m,n\}!}{\max\{m,n\}!}}\left(\pm 2i\eta\right)^{|n-m|} L_{\min\{m,n\}}^{|n-m|}\left(4\eta^2\right), \tag{4}$$

in terms of the associated Laguerre polynomial $L_a^{(b)}(x)$. This hopping exhibits the asymptotic behaviours of $l_{nm}^\pm(\eta\to 0) = \delta_{mn}$ and $l_{nm}^\pm(\eta\to\infty) = 0$ (End Matter D).

Owing to weighted hopping $l_{nm}^\pm = (l_{mn}^\mp)^*$ that spans all orders of long-range hopping without Bloch boundary conditions, the model of Eq. (3) must be interpreted as a complex magnetic graph[22-25] rather than as a lattice. The QRM forms a bipartite graph composed of two disjoint semi-infinite sets, represented by the state vectors $\Psi_\pm = [c_1^\pm, c_2^\pm, \ldots]^\mathrm{T}$, while the self-loops determined by onsite energies vary linearly as $n\hbar\omega_0$ with additional perturbations of $\pm|\mathbf{h}_\parallel|$.

To characterize the connectivity of the magnetic graph, we apply a gauge transformation that flattens the linear variation of the self-loops, which leads to (End Matter E):

$$i\hbar\frac{\partial}{\partial t}\begin{bmatrix}\Psi_+\\ \Psi_-\end{bmatrix} = \begin{bmatrix}+|\mathbf{h}_\parallel|I & K_\pm'\\ K_\mp' & -|\mathbf{h}_\parallel|I\end{bmatrix}\begin{bmatrix}\Psi_+\\ \Psi_-\end{bmatrix} \triangleq \Omega'\begin{bmatrix}\Psi_+\\ \Psi_-\end{bmatrix}, \tag{5}$$

where $[K_\pm']_{mn} = \exp[i(m-n)\omega_0 t][K_\pm]_{mn}$ denotes the element of the transformed hopping matrix with $[K_\pm]_{mn} = |\mathbf{h}_\perp| l_{mn}^\pm$ (Fig. 1d). Because $|l_{mn}^\pm| \to 0$ for sufficiently large $|m-n|$ (End Matter D), we can introduce a coupling-strength-dependent maximum hopping order $\Delta N_\eta$ such that $l_{mn}^\pm \approx 0$ for $|m-n| > \Delta N_\eta$. This condition implies a finite temporal periodicity of $2\pi/[\omega_0\Delta N_\eta]$, yielding a Floquet graph composed of identical nodes except for the perturbations $\pm|\mathbf{h}_\parallel|$ on the disjoint states $\Psi_{\pm}$. We refer to the resulting single-atom QED graph as the Floquet Rabi (FR) graph, denoted by $G_{\mathrm{FR}} = (V_{\mathrm{FR}}, E_{\mathrm{FR}})$, where $V_{\mathrm{FR}}$ represents the set of quantum states $|\varphi_{\pm}\rangle|n\rangle$ and $E_{\mathrm{FR}}$ denotes the set of edges characterized by $K_{\pm}'$. Truncating the bosonic Hilbert space at $n = N_{\max}$ renders the FR graph finite.

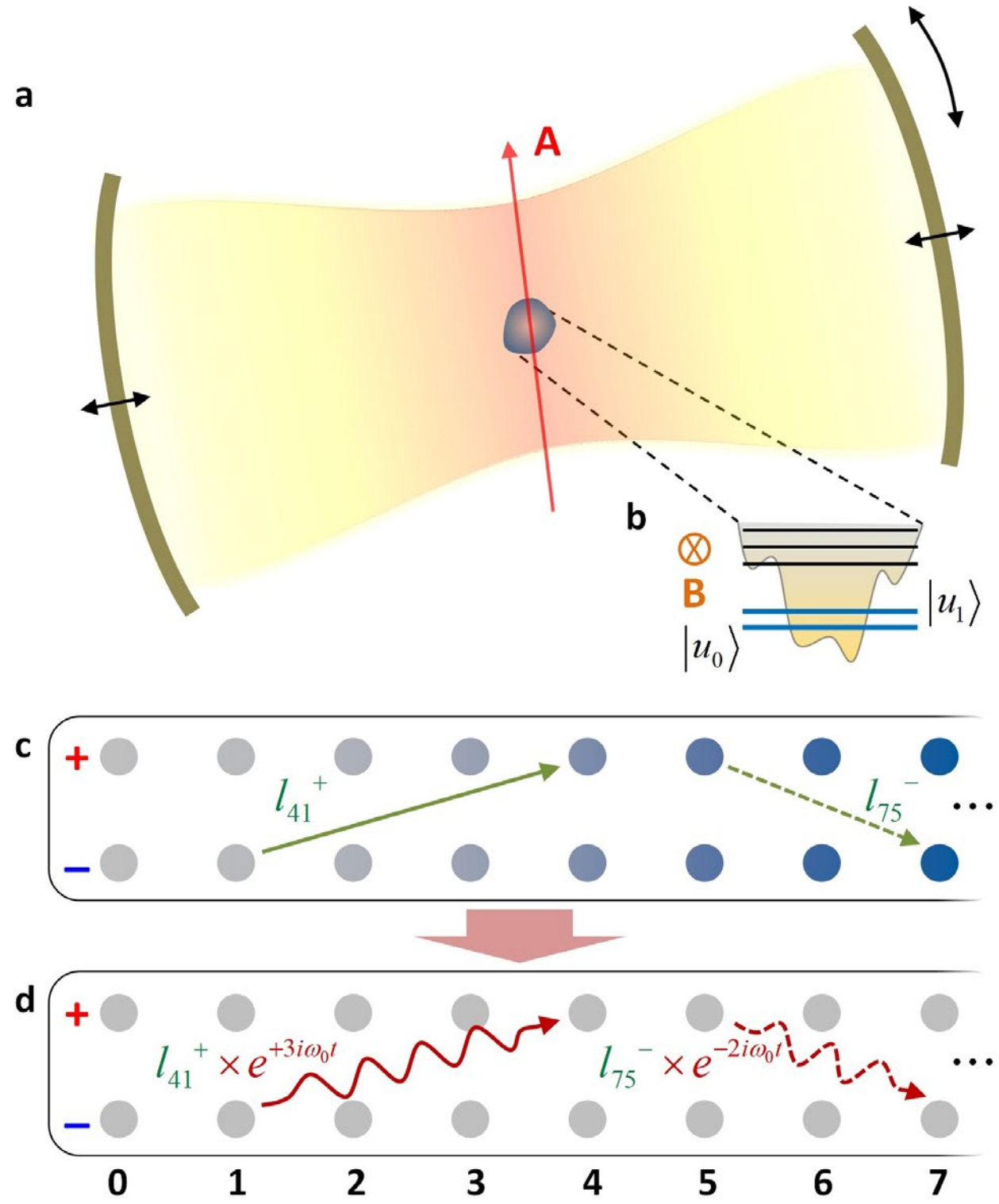

**FIG. 1. Magnetic graph model for single-atom cavity QED. a,b,** Schematics for the generalized QRM (**a**) and the eigenstates of an arbitrary atom (**b**). Black arrows in **a** indicate cavity modulations. **c,** Graph model of Eq. (4), with linearly varying self-loops indicated by the graded colour of the nodes (circles). **d,** The FR graph obtained via a local gauge transformation of the graph in **c**.

Figure 2 shows the FR graph connectivity depending on $\eta$. According to Eq. (4), the graph edges are classified by their hopping phases of $\gamma\pi/2$ ($\gamma = 0, 1, 2,$ and $3$), determined by the hopping order with the sign of the associated Laguerre polynomial. Although the nearest-neighbour (NN) edges are dominant in the weak-coupling regime (Fig. 2a), the effect of higher-order hopping increases with enhanced coupling (Fig. 2b), eventually comprising highly complicated mixing of complex-valued weights over the entire graph (Fig. 2c,d).

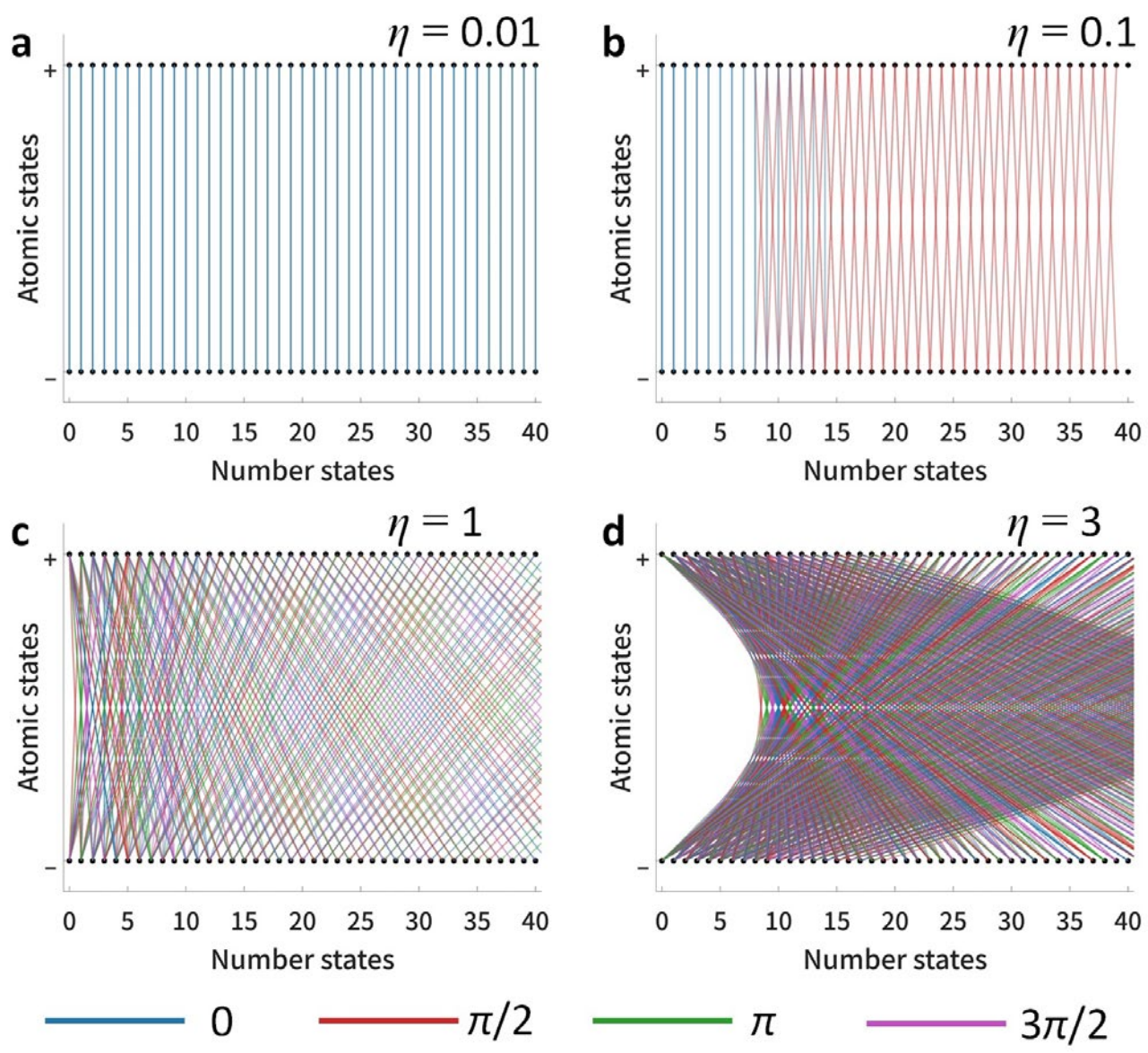

**FIG. 2. Graph connectivity for different values of *η* at *t* = 0.** Edges with weights greater than 50 % of the global maximum are shown for visibility, and the strength of each weight is indicated by transparency. To suppress errors arising from the finite-dimensional truncation of the Fock space, we employ a large cutoff for the maximum number state: $|n = N_{\max} = 200\rangle$.

## Magnetic Laplacian

To examine the connectivity of the FR graph—particularly in the USC and DSC regimes—we employ the normalized magnetic Laplacian[25], while setting $|\mathbf{h}_{\parallel}| = 0$. For the magnetic graph operator $\Omega'$ in Eq. (5), its Laplacian is $L' = I - D'^{-1/2}\Omega' D'^{-1/2}$, where $D'$ is the diagonal degree matrix constructed from the elementwise absolute value matrix $|\Omega'|$. Although $L' = L'(t)$ has a Floquet form owing to $K_{\pm}'$, it is readily proved that the eigenspectrum of $L'$ remains invariant under this temporal variation (End Matter F), indicating that the connectivity of the FR graph is preserved. Therefore, we investigate the graph connectivity using the static magnetic Laplacian $L = I - D^{-1/2}\Omega D^{-1/2}$ with

$$\Omega = \begin{bmatrix} 0 & K_{\pm} \\ K_{\mp} & 0 \end{bmatrix}, \tag{6}$$

where $D$ is the diagonal degree matrix of $|\Omega|$.

Using the Laplacian, we solve the eigenvalue problem $Lv_m = \lambda_m v_m$ for positive integers $m$ and $\lambda_m \leq \lambda_{m+1}$. According to the generalized Cheeger's inequality[25], the critical metric for characterizing graph connectivity is the first eigenvalue $0 \leq \lambda_1 \leq 1$ (End Matter G), which provides a lower bound on the cost of separating a large phase-coherent (or magnetic-flux-free) subgraph $G_S \subseteq G_{FR}$. To illustrate this analysis, consider the example graph in Fig. 3a, possessing three subgraphs: green, purple, and red ones. In terms of phases and strengths of edge weights, the red subgraph corresponds to a phase-coherent—that is, the absence of a magnetic flux—region ('0') with a low separation cost due to the weight bottleneck (orange arrows). By contrast, the green subgraph exhibits a nonzero magnetic flux—referred to as phase frustration in graph theory[25]—that must be avoided for coherence, while the purple one is strongly connected to the remainder of the graph, and therefore, enforces a high separation cost. Therefore, $\lambda_1$ represents the minimum cost of separating the red subgraph from the graph.

Figure 3b shows the variation of $\lambda_1$ with $\eta$. Over the weak- to strong-coupling range ($\eta < 10^{-1}$), the plot simply follows the linear law $\lambda_1 = 2\eta$. Within the interval of $0 \leq \lambda_1 \leq 1$, the regimes beyond the USC are characterized by substantial deviations from the linearity, whereas the DSC regime is identified by the saturation of $\lambda_1$ at unity. Consequently, our FR graph enables a $\lambda_1$-based classification of single-atom cavity QED, which continuously quantifies the cost of separating a nonmagnetic subgraph. We note that $\lambda_1$ provides a single, continuous-valued metric derived directly from the connectivity of quantum states in the generalized QRM, avoiding the need to invoke regime-specific approximation conditions[13].

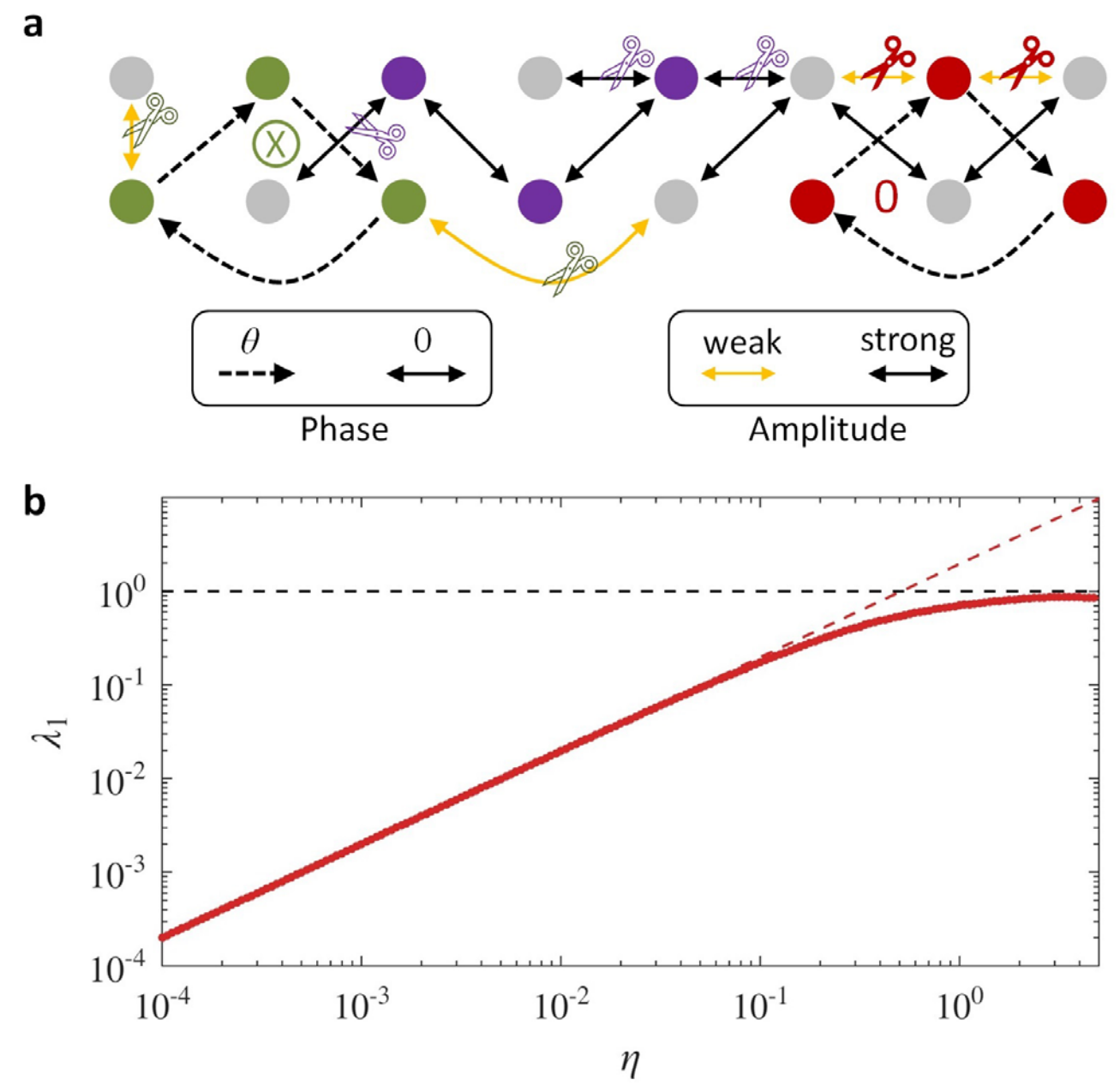

**FIG. 3. Stronger coupling indicates a higher cost for separating a large nonmagnetic subgraph. a,** Schematic of the generalized Cheeger's inequality. Sets of green, purple, and red circles denote subgraphs. Scissors mark cuts—setting the edge weight to zero—used to separate subgraphs. Directional dashed arrows and double-headed solid arrows indicate hopping with nonzero and zero phase evolution, respectively. Black and orange arrows denote strong and weak hopping strengths, respectively. **b,** $\lambda_1$ as a function of $\eta$. The red and black dashed lines denote the weak-coupling asymptote and the upper bound on $\lambda_1$, respectively.

## Conductance and frustration

Although $\lambda_1$ offers a single-parameter classification of single-atom cavity QED, an increase in $\lambda_1$ with stronger coupling can originate from two distinct mechanisms: (i) a high separation cost due to the alleviation of edge-weight bottlenecks, and (ii) magnetic-flux-induced phase frustration. Interestingly, their impacts on quantum states, specifically in terms of transport over the FR graph, are opposite: mitigating bottlenecks enhances transport—that is, higher conductance[26]—whereas the frustration with phase incoherence, leads to localization.

To identify the origin of the variation in $\lambda_1$, we disentangle the two contributions by randomly constructing subgraphs and examining their closed loops. Owing to the bipartite structure of the FR graph, each closed loop contains an equal number of $|\varphi_+\rangle|n\rangle$ and $|\varphi_-\rangle|n\rangle$ site vertices (Fig. 4a,b). Accordingly, we form a subgraph, $G_S{}^q = (V_S{}^q, E_S{}^q)$, by randomly selecting $q$ pairs across the bipartition, as $V_S{}^q = \{(|r_+\rangle, |s_-\rangle)|r, s = 1, 2, \dots, q\}$, where $|r_\pm\rangle \coloneqq |\varphi_\pm\rangle|n(r_\pm)\rangle$, and $n(r) \sim \mathrm{Unif}\{0, 1, 2, \dots, N_{\max}\}$ denotes uniform sampling over the number states. For $q > 2$, a closed loop can be defined as $e(1_-, 1_+) \to e(2_+, 1_-) \to e(2_-, 2_+) \to \dots \to e(1_+, q_-)$, where $e(r_\sigma, s_\tau) \in E_S{}^q$ is the directed edge from $|s_\tau\rangle$ to $|r_\sigma\rangle$ with complex weight $w(r_\sigma, s_\tau)$ for $\sigma, \tau = (\pm)$.

For each subgraph $G_S{}^q$ composed of $2q$ nodes and one of its closed loops $\Gamma_q$, we evaluate the origin of the $\lambda_1$ variation using the conductance and the loop weight:

$$C_q = \frac{\sum_{r_\sigma \in \left(V_S{}^q\right)^{\mathrm{C}}, s_\tau \in V_S{}^q} \left|w(r_\sigma, s_\tau)\right|}{\sum_{r_\sigma \in V_S{}^q, s_\tau \in V_{\mathrm{FR}}} \left|w(r_\sigma, s_\tau)\right|}, \quad W_q = \prod_{e(r_\sigma, s_\tau) \in \Gamma_q} w(r_\sigma, s_\tau). \tag{7}$$

where their $q$-dependencies characterize the degree of globality of these two measures. The conductance $C_q$ measures the total strength of edges connecting $G_S{}^q$ to the rest of the graph, normalized by the volume of $G_S{}^q$—the sum of node degrees. This quantity characterizes the separation cost of $G_S{}^q$, by probing the presence of bottlenecks[26].

The magnetic signature of the FR graph is quantified by $W_q = |W_q|\exp(i\Phi_q)$. Its phase $\Phi_q$ is a topological invariant quantized to 0 (trivial) or $\pi$ (nontrivial) modulo $2\pi$, which characterizes the parity conversion of the number states along the loop (End Matter H). Consequently, a large fraction of subgraphs having nontrivial phase $\Phi_q = \pi$ leads to destructive interference and the resulting phase frustration, thereby contributing to increasing $\lambda_1$. Because the FR graph is weighted, this fraction is weighted by $|W_q|$.

Figure 4c shows the ensemble-averaged conductance $\langle C_q \rangle$ over random subgraphs. The result demonstrates that conductance does not contribute to the $\lambda_1$ variation, as $\langle C_q \rangle$ remains nearly constant with respect to $\eta$; that is, the fraction of bottlenecks reflecting the graph structure is independent of $\eta$. Moreover, the approximately linear dependence of $\langle C_q \rangle$ on the subgraph size $2q$ indicates that the FR graph is quasi-one-dimensional, reflecting the dimensional distinction between $|\varphi_\pm\rangle$ and $|n\rangle$.

Consequently, the coupling-induced change in graph connectivity originates solely from phase frustration. To verify this prediction, we evaluate $W_q$. In the weak and strong coupling regimes, the graph connectivity is dominated by NN hopping, hindering the closed-loop formation. As $\eta$ increases, the system progressively enables higher $q$-loop formation, reaching > 80% for $q$ = 2 and > 40% for $q$ = 10 as it approaches the DSC regime (Supplementary Note S1).

Using the loop formation, we compute the ensemble-averaged $|W_q|^{1/(2q)}$—the averaged connection strength (End Matter I)—separately for nontrivial ($\Phi_q = \pi$) and trivial ($\Phi_q = 0$) loops (Fig. 4d). In parallel, we evaluate the eigenstates of $H_\mathrm{C}|\psi_n\rangle = E_n|\psi_n\rangle$ via two measures: the inverse participation ratio (IPR) $\mathrm{IPR}_n = \sum_{v=1}^{2(N_\mathrm{max}+1)} \left|\langle v|\psi_n\rangle\right|^4$ which quantifies the state localization (Fig. 4e) and the eigenvalues $E_n$ (Fig. 4f).

The results demonstrate that phase frustration governs the coupling-induced state transition in cavity QED. As shown in Fig. 4d, approximately half of the formed loops are nontrivial. The destructive interference induced by these nontrivial loops leads to localization (Fig. 4e) and the consequent spectral degeneracies (Fig. 4f). While the weak- and strong-coupling regimes are dominated by $q$ = 2 loops, the USC regime is characterized by the mixed emergence of higher-order loops ($q$ > 2). We note that $|W_q|^{1/(2q)}$ saturates as the system approaches the DSC regime across different $q$ values, indicating that phase frustration becomes uniform across the graph, leading to similar values of $\langle\Phi_q\rangle$ regardless of $q$ or loop configuration.

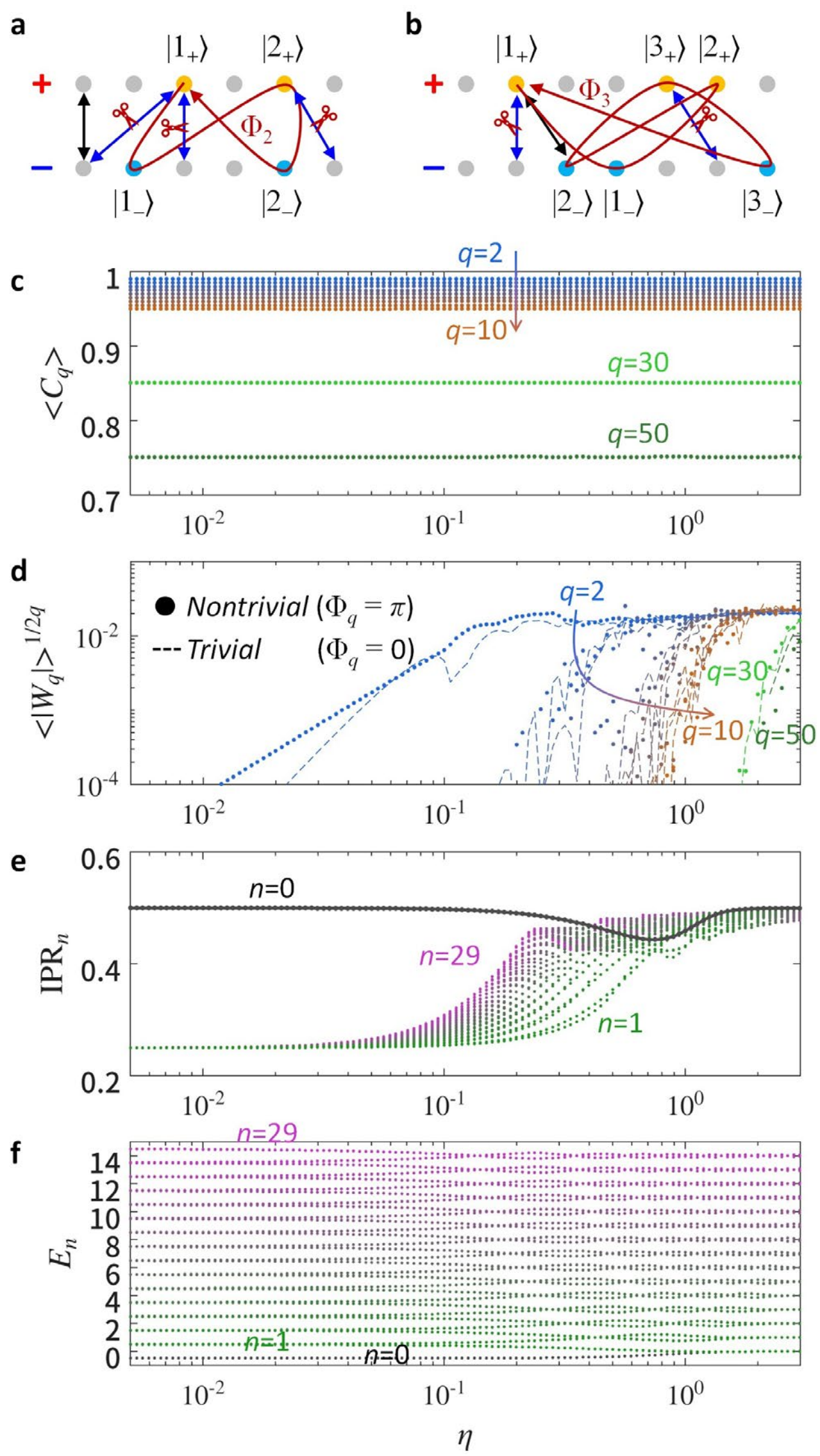

**FIG. 4. Phase frustration governs the USC. a,b,** Schematics of subgraphs (coloured nodes) illustrating closed loops for $q = 2$ (**a**) and $q = 3$ (**b**). Black and blue arrows indicate example edges, while the blue ones must be cut to separate the subgraphs. Red curved arrows denote closed loops. **c,d,** Graph measures: $\langle C_q\rangle$ (**c**) and $|W_q|^{1/(2q)}$ (**d**), as functions of $\eta$, for different $q$ values. Both measures are computed over $2\times10^4$ random subgraphs, while $|W_q|^{1/(2q)}$ is evaluated separately for nontrivial (dots) and trivial (dashed lines) phases. **e,f,** Eigenstate analysis: IPR (**e**) and $E_n$ (**f**) as functions of $\eta$. The first 30 eigenstates are plotted.

## Discussion

Although we apply the magnetic graph modelling to single-atom cavity QED, the approach can be readily extended to many-body systems, such as the Dicke and Rabi-Hubbard models. Using the Coulomb-gauge models developed under the gauge principle[1,14], the emergence of long-range hopping in the Fock-state-lattice can be transformed into a large-scale magnetic graph with identical nodes under Floquet boundary conditions. This extension enables the analysis of large-scale quantum phenomena using subgraph separation and phase frustration. Therefore, our graph model can operate as a general framework for classifying cavity QED including many-body phenomena.

In conclusion, we investigated complex graph dynamics in the simplest quantum system. For the generalized QRM that incorporates vector gauge fields and cavity modulations, we constructed its magnetic graph representation. We quantified graph connectivity through $\lambda_1$, which indicates the USC and DSC regimes. We analysed the origin of $\lambda_1$ in terms of separation cost and phase frustration, demonstrating the critical role of frustration. This approach provides a universal framework for interpreting atom-light interactions as a complex graph, enabling the characterization of quantum-state properties, such as robustness from localization on the graph. The application of graph-based methods, including spectral clustering and graph neural networks, would be promising future research.

## Methods

### A. Position operator and dipole moments

Following the procedures in previous works[1,14], we sequentially apply truncation and minimal coupling replacement to the Schrödinger equation in the infinite-dimensional Hilbert space $\mathcal{H}$. As a preliminary step, we prepare projection-based representation of the position operator and the dipole moments. Consider an arbitrary orthonormal basis $\{|u_n\rangle | n = 0, 1, \ldots, \infty\}$ of $\mathcal{H}$. To express the position operator, we introduce the projection operator

$$P = \sum_{r \in N_\mathrm{T}} |u_r\rangle\langle u_r|, \tag{A1}$$

which leads to $\mathbf{x} = P\mathbf{x}P$ when $N_\mathrm{T} = \{0, 1, \ldots, \infty\}$, and thus, $P = I$. The position operator is represented as

$$\mathbf{x} = \sum_{r,s \in N_\mathrm{T}} \frac{\mathbf{d}_{rs}}{q} |u_r\rangle\langle u_s|, \tag{A2}$$

where $\mathbf{d}_{rs} = q\langle u_r|\mathbf{x}|u_s\rangle$ corresponds to the vector dipole matrix element between the $r$-th and $s$-th basis vectors.

**B. Generalized QRM**

To analyse a single-atom cavity QED, the infinite-dimensional Hilbert space $\mathcal{H}$ needs to be truncated to a two-level subspace $\mathcal{H}_\mathrm{T}$. This corresponds to introducing a truncated index set $N_\mathrm{T}$ = $\{0, 1\}$ in Eqs. (A1) and (A2), indicating two states $\{|u_0\rangle, |u_1\rangle\}$ that yield physics well separated from that of the remaining states (Fig. 1b). An arbitrary bare Hamiltonian $H_0$ on $\mathcal{H}_\mathrm{T}$ then becomes $H_0 = h_0 I + \mathbf{h}\cdot\boldsymbol{\sigma}$. When $|u_0\rangle$ and $|u_1\rangle$ are spatially localized, the dipole moment vectors are also well defined in $\mathcal{H}_\mathrm{T}$: the averaged and differential permanent dipole moments $\mathbf{d}_\pm = (\mathbf{d}_{11} \pm \mathbf{d}_{00})/2$ and the transition dipole moment $\mathbf{d} = \mathbf{d}_\mathrm{r} + i\mathbf{d}_\mathrm{i}$ with $\mathbf{d}_\mathrm{r} = \mathrm{Re}[\mathbf{d}_{10}]$ and $\mathbf{d}_\mathrm{i} = \mathrm{Im}[\mathbf{d}_{10}]$.

According to minimal coupling replacement with a single-mode gauge field $\mathbf{A} = \mathbf{A}_0(a^\dagger + a)$ under the dipole approximation, the gauge-invariant Hamiltonian in the Coulomb gauge is obtained as $H_C = UH_0U^\dagger + \hbar\omega_0 a^\dagger a^1$, where $\omega_0$ denotes the frequency of light, and $U = \exp[(i/\hbar)q\mathbf{A}\cdot\mathbf{x}]$ is the unitary operator on $\mathcal{H}_T$. Using the truncated position operator, $\mathbf{x} = (1/q)(\mathbf{d}_+ I + \mathbf{d}_r\sigma_x - \mathbf{d}_i\sigma_y + \mathbf{d}_-\sigma_z)$, we obtain $U = \exp[(i/\hbar)(g_0 I + \mathbf{g}\cdot\boldsymbol{\sigma})(a^\dagger + a)]$, where $g_0 = \mathbf{A}_0\cdot\mathbf{d}_+$ and $\mathbf{g} = (\mathbf{A}_0\cdot\mathbf{d}_r)\mathbf{i} - (\mathbf{A}_0\cdot\mathbf{d}_i)\mathbf{j} + (\mathbf{A}_0\cdot\mathbf{d}_-)\mathbf{k}$.

To obtain $H_C = UH_0U^\dagger + \hbar\omega_0 a^\dagger a$, we employ the BCH formula[27]. When setting $X = (i/\hbar)(g_0 I + \mathbf{g}\cdot\boldsymbol{\sigma})(a^\dagger + a)$ for $U = \exp[(i/\hbar)(g_0 I + \mathbf{g}\cdot\boldsymbol{\sigma})(a^\dagger + a)]$, and $Y = H_0 = h_0 I + \mathbf{h}\cdot\boldsymbol{\sigma}$, their commutation gives

$$[X,Y] = \frac{2}{\hbar}\left(a^\dagger + a\right)\left(\mathbf{h}\times\mathbf{g}\right)\cdot\boldsymbol{\sigma}, \tag{B1}$$

Through the recursive calculation, we obtain

$$[X,Y]_m = \left(\frac{2}{\hbar}\right)^m \left(a^\dagger + a\right)^m T_\mathbf{g}^{\,m}(\mathbf{h})\cdot\boldsymbol{\sigma}, \tag{B2}$$

where $[X,Y]_m$ denotes a series of the commutation satisfying $[X,Y]_m = [X,[X,Y]_{m-1}]$ with $[X,Y]_1 = [X,Y]$, and $T_\mathbf{g}^m(\mathbf{h})$ denotes a series of the cross product as $T_\mathbf{g}^m(\mathbf{h}) = T_\mathbf{g}^{m-1}(\mathbf{h}) \times \mathbf{g}$ and $T_\mathbf{g}^0(\mathbf{h}) = \mathbf{h}$. Therefore, the BCH formula for $e^X Y e^{-X}$ leads to the following gauge-invariant Hamiltonian:

$$\begin{aligned} H_C &= \hbar\omega_0 a^\dagger a + h_0 I + \sum_{m=0}^{\infty}\frac{1}{m!}\left(\frac{2}{\hbar}\right)^m \left(a^\dagger + a\right)^m T_\mathbf{g}^{\,m}(\mathbf{h})\cdot\boldsymbol{\sigma} \\ &\triangleq \hbar\omega_0 a^\dagger a + h_0 I + \left[\exp\left(2\frac{a^\dagger + a}{\hbar}T_\mathbf{g}\right)\mathbf{h}\right]\cdot\boldsymbol{\sigma}. \end{aligned} \tag{B3}$$

By applying the Rodrigues rotation formula[28], we obtain

$$\begin{aligned} H_C = \hbar\omega_0 a^\dagger a + h_0 I + \Bigg[ &\mathbf{h}_\parallel + \mathbf{h}_\perp \cos\left(\frac{2(a^\dagger + a)}{\hbar}|\mathbf{g}|\right) \\ &+ \mathbf{h}\times\frac{\mathbf{g}}{|\mathbf{g}|}\sin\left(\frac{2(a^\dagger + a)}{\hbar}|\mathbf{g}|\right)\Bigg]\cdot\boldsymbol{\sigma}, \end{aligned} \tag{B4}$$

where $\mathbf{h}_\parallel$ and $\mathbf{h}_\perp$ denote the parallel and perpendicular components of $\mathbf{h}$ with respect to $\mathbf{g}$, respectively, as $\mathbf{h}_\parallel = (\mathbf{h}\cdot\mathbf{g})\mathbf{g}/|\mathbf{g}|^2$ and $\mathbf{h}_\perp = \mathbf{h} - \mathbf{h}_\parallel$.

Using the displacement operator[29], $D(\alpha) = \exp(\alpha a^\dagger - \alpha^* a)$, the trigonometric operators in Eq. (B4) can be expressed in terms of $D$, as $\cos[\alpha(a^\dagger + a)] = [D(+i\alpha) + D(-i\alpha)]/2$ and $\sin[\alpha(a^\dagger + a)] = [D(+i\alpha) - D(-i\alpha)]/2i$. By substituting these relations into Eq. (B4), we obtain Eqs. (1) and (2).

Equation (1) is independent of the chosen basis $\{|u_0\rangle, |u_1\rangle\}$, allowing one to treat an arbitrary Hamiltonian $H_0$ characterized by $\mathbf{h}$, together with the generalized vector dipole moments, $\mathbf{d}_\pm$ and $\mathbf{d}$. We note that $\mathbf{d}_-$ and $\mathbf{d}_\mathrm{i}$ can be set to zero by choosing $\{|u_0\rangle, |u_1\rangle\}$ to satisfy parity and time-reversal symmetries. In this case, the system is governed by the Hamiltonian vector $\mathbf{h}$ and the simplified vector $\mathbf{g} = (\mathbf{A}_0\cdot\mathbf{d}_\mathrm{r})\mathbf{I}$, which depends solely on the vector field $\mathbf{A}_0$.

However, for consistency with experimental measurements, the eigenbasis $\{|u_0\rangle, |u_1\rangle\}$ defined by $H_0|u_n\rangle = E_n|u_n\rangle$ is conventionally employed. Under the two-level truncation, we obtain $H_0 = [(E_1 + E_0)/2]I + [(E_1 - E_0)/2]\sigma_z$, leading to $\mathbf{h} = (E_{10}/2)\mathbf{k}$ for $E_{10} = E_1 - E_0$. In this representation, the vector $\mathbf{g}$ encodes the coupling between dipole moments and light.

## C. Fock-state lattice representation

When applying $H_\mathrm{C}$ in Eq. (1) to the ansatz $|\psi\rangle = \sum_n (c_n^+|\varphi_+\rangle + c_n^-|\varphi_-\rangle)|n\rangle$, we obtain

$$\begin{aligned} H_\mathrm{C}|\psi\rangle = \sum_m & \left(m\hbar\omega_0 + h_0\right)\left(c_m^+|\varphi_+\rangle + c_m^-|\varphi_-\rangle\right)|m\rangle \\ & + |\mathbf{h}_\parallel|\left(c_m^+|\varphi_+\rangle - c_m^-|\varphi_-\rangle\right)|m\rangle \\ & + |\mathbf{h}_\perp|\left(c_m^-|\varphi_+\rangle\right)D(+2i\eta)|m\rangle \\ & + |\mathbf{h}_\perp|\left(c_m^+|\varphi_-\rangle\right)D(-2i\eta)|m\rangle, \end{aligned} \tag{C1}$$

using the relations: $(\hat{\mathbf{h}}_\parallel\cdot\boldsymbol{\sigma})|\varphi_\pm\rangle = \pm|\varphi_\pm\rangle$, $(\hat{\mathbf{h}}_\pm\cdot\boldsymbol{\sigma})|\varphi_\pm\rangle = O$, and $(\hat{\mathbf{h}}_\mp\cdot\boldsymbol{\sigma})|\varphi_\pm\rangle = \pm|\varphi_\mp\rangle$. By exerting $\langle\varphi_+|\langle n|$ and $\langle\varphi_-|\langle n|$ on the right side of Eq. (C1), we obtain Eq. (3) with $l_{nm}^\pm = \langle n|D(\pm 2i\eta)|m\rangle$.

## D. Asymptotic behaviours of hopping

In analysing the hopping coefficient, $l_{nm}^{\pm}$ in Eq. (4), we introduce the coefficients $s \coloneqq \min\{m,n\}$ and $\Delta \coloneqq |m-n|$ for simplicity, which lead to

$$l_{nm}^{\pm} = e^{-2\eta^2} \sqrt{\frac{s!}{(s+\Delta)!}} \left(\pm 2i\eta\right)^{\Delta} L_s^{\Delta}\left(4\eta^2\right). \tag{D1}$$

The associated Laguerre polynomial has the closed form of

$$L_s^{\Delta}\left(4\eta^2\right) = \sum_{k=0}^{s} (-1)^k \binom{s+\Delta}{s-k} \frac{\left(4\eta^2\right)^k}{k!}. \tag{D2}$$

When $\eta \to 0$, we use $\exp(-2\hbar\eta^2) = 1 - 2\hbar\eta^2 + O(\eta^4)$, achieving the following asymptotic behaviour:

$$\begin{aligned} l_{nm}^{\pm} = {} & \frac{\left(\pm 2i\eta\right)^{\Delta}}{\Delta!} \sqrt{\frac{(s+\Delta)!}{s!}} \left[1 - 2\left(1 + \frac{2s}{\Delta+1}\right)\eta^2\right] \\ & + O\left(\eta^4\right), \end{aligned} \tag{D3}$$

which gives $l_{nn}^{\pm} = 1 - 2(1+2n)\eta^2 + O(\eta^4)$. Therefore, in the weak coupling regime, the $(\pm 2i\eta)^{\Delta}$ term in Eq. (D3) indicates that the NN hopping between spin states $l_{nn}^{\pm}$ dominates the QRM dynamics.

When $\eta \to \infty$, the higher-order terms in Eq. (D2) dominate, as $L_s^{\Delta}(4\eta^2) = (-1)^s(4\eta^2)^s[1 - s(s+\Delta)/(4\eta^2)]/s! + O(\eta^{2s-4})$, which leads to

$$\begin{aligned} l_{nm}^{\pm} = {} & \frac{(\pm i)^{\Delta}(-1)^s}{\sqrt{s!(s+\Delta)!}} (2\eta)^{m+n} e^{-2\eta^2} \\ & \times \left[1 - \frac{s(s+\Delta)}{4\eta^2} + O\left(\eta^{-4}\right)\right]. \end{aligned} \tag{D4}$$

Therefore, in the DSC regime with $\eta >> 1$, $l_{nm}^{\pm}$ decays as the Gaussian, $\exp(-2\eta^2)$, multiplied by the polynomial factor $\eta^{m+n}$. For a given large value of $\eta$, the increase of $s = \min\{m,n\}$ or $\Delta = |m-n|$ leads to $|l_{nm}^{\pm}| \to 0$.

## E. Transformation to FR graph

For $\Psi_\pm = [c_1^\pm, c_2^\pm, \ldots]^T$, Eq. (3) can be rewritten as

$$i\hbar\frac{\partial}{\partial t}\begin{bmatrix}\Psi_+\\ \Psi_-\end{bmatrix}=\begin{bmatrix}B+\left|\mathbf{h}_\parallel\right|I & K_\pm\\ K_\mp & B-\left|\mathbf{h}_\parallel\right|I\end{bmatrix}\begin{bmatrix}\Psi_+\\ \Psi_-\end{bmatrix}, \tag{E1}$$

where $K_\pm$ denote the hopping matrices with $[K_\pm]_{mn} = |\mathbf{h}_\perp| l_{mn}^\pm$, and $B$ is the diagonal matrix given by $[B]_{nn} = n\hbar\omega_0 + h_0$. To apply the analytical tool for the connectivity of magnetic graphs[22-25], we need to eliminate $B \pm |\mathbf{h}_\parallel| I$ in Eq. (E1). The term $|\mathbf{h}_\parallel|$ can be treated as a perturbation, because it vanishes when the atomic potential satisfies parity symmetry with $|\mathbf{d}_\perp| = 0$. To remove the remaining terms, we apply the following gauge transformation analogous to the standard treatment of Wannier-Stark ladders and Bloch oscillations[30,31]:

$$\begin{aligned}i\hbar\frac{\partial}{\partial t}\begin{bmatrix}\Psi_+'\\ \Psi_-'\end{bmatrix}&=U\begin{bmatrix}B+\left|\mathbf{h}_\parallel\right|I & K_\pm\\ K_\mp & B-\left|\mathbf{h}_\parallel\right|I\end{bmatrix}U^\dagger\begin{bmatrix}\Psi_+'\\ \Psi_-'\end{bmatrix}\\ &+i\hbar\left(\frac{\partial U}{\partial t}\right)U^\dagger\begin{bmatrix}\Psi_+'\\ \Psi_-'\end{bmatrix},\end{aligned} \tag{E2}$$

where the transformation is defined by

$$U=\begin{bmatrix}V & O\\ O & V\end{bmatrix},\quad V=\exp\left[\frac{i}{\hbar}\sum_n\left(n\hbar\omega_0+h_0\right)\left|n\right\rangle\left\langle n\right|\right]. \tag{E3}$$

We then obtain

$$i\hbar\frac{\partial}{\partial t}\begin{bmatrix}\Psi_+'\\ \Psi_-'\end{bmatrix}=\begin{bmatrix}+\left|\mathbf{h}_\parallel\right|I & UK_\pm U^\dagger\\ UK_\mp U^\dagger & -\left|\mathbf{h}_\parallel\right|I\end{bmatrix}\begin{bmatrix}\Psi_+'\\ \Psi_-'\end{bmatrix}, \tag{E4}$$

which becomes Eq. (5) upon redefining $\Psi_\pm \coloneqq \Psi_\pm'$.

## F. Magnetic Laplacian analysis for FR graph

Consider the Floquet magnetic Laplacian of the FR graph: $L'(t) = I - D'^{-1/2}\Omega' D'^{-1/2}$, where

$$\Omega' = \begin{bmatrix} 0 & K_{\pm}' \\ K_{\mp}' & 0 \end{bmatrix} = V \begin{bmatrix} 0 & K_{\pm} \\ K_{\mp} & 0 \end{bmatrix} V^{\dagger}, \tag{F1}$$

for the diagonal matrix $V$ in Eq. (E3). Therefore, we can obtain $L'(t) = I - VD'^{-1/2}\Omega' D'^{-1/2}V^{\dagger}$ for diagonal matrices $V$ and $D'$. We also note that $|\Omega'| = |\Omega|$ because of the relationship $[K_{\pm}']_{mn} = \exp[i(m - n)\omega_0 t][K_{\pm}]_{mn}$, which leads to $D = D'$. Therefore, the relationship between the Floquet magnetic Laplacian and the static magnetic Laplacian becomes

$$L = V^{\dagger} L'(t) V. \tag{F2}$$

Because a unitary transformation preserves the eigenspectrum, we can analyse the connectivity of the FR graph using $L$.

## G. Bounds on $\lambda_1$

Because $L$ is positive semidefinite, its smallest eigenvalue satisfies $\lambda_1 \geq 0$. We also note that the sum of the eigenvalues equals Tr[$L$]. Using the cyclicity of the trace, we obtain

$$\mathrm{Tr}\left[L\right] = \mathrm{Tr}\left[I\right] - \mathrm{Tr}\left[\Omega D^{-1}\right] = N, \tag{G1}$$

because $\mathrm{Tr}[\Omega D^{-1}] = 0$ with $[\Omega]_{nn} = 0$ and the diagonal matrix $D$. The average of the eigenvalues is then 1. Because $\lambda_1$ is no greater than this average for the positive semidefinite $L$, we obtain $0 \leq \lambda_1 \leq 1$.

## H. Topological magnetic fluxes

Consider a closed loop, $e(1_-, 1_+) \rightarrow e(2_+, 1_-) \rightarrow e(2_-, 2_+) \rightarrow \ldots \rightarrow e(1_+, q_-)$, with the corresponding complex edge weights $w(r_\sigma, s_\tau)$ for $\sigma, \tau = (\pm)$. According to Eqs. (4) and (6), the weight $w(r_\sigma, s_\tau)$ can have a hopping phase $\gamma\pi/2$ ($\gamma = 0, 1, 2,$ and 3), where even and odd values of $\gamma$ correspond to even and odd values of $|n(r_\sigma) - n(s_\tau)|$, respectively. Therefore, the number of edges with odd $\gamma$ can be interpreted as counting the number of parity conversions along the sequence $\{n(1_+), n(1_-), n(2_+), n(2_-), \ldots, n(q_+), n(q_-), n(1_+)\}$. Because the sequence starts and ends with $n(1_+)$, which has the same parity, the number of parity conversions must be even. Therefore, the accumulated phase $\Phi_q$ is quantized to 0 (trivial) or $\pi$ (nontrivial). The quantized value reflects the topological nature of the weighted loop defined by the sequence $\{n(1_+), n(1_-), n(2_+), n(2_-), \ldots, n(q_+), n(q_-), n(1_+)\}$.

### I. Connection strengths

According to Eq. (7), $W_q$ is defined as the multiplication of $2q$ weights $w(r_\sigma, s_\tau)$. When any of these weights is zero, $W_q$ becomes zero, indicating a disconnected loop. Otherwise, $|W_q|^{1/(2q)}$ quantifies the average connection strength per edge along the loop. This quantity allows for comparing loops with different values of $q$, including disconnected ones.

## Data availability

The data used in this study are available from the corresponding authors upon request.

## Code availability

The codes used in this study are available from the corresponding authors upon request.

## Acknowledgements

We acknowledge financial support from the National Research Foundation of Korea (NRF) through the Basic Research Laboratory (No. RS-2024-00397664), Innovation Research Center (No. RS-2024-00413957), Pilot and Feasibility Grants (No. RS-2025-19912971), Young

Researcher Program (No. RS-2025-00552989), and Midcareer Researcher Program (No. RS-2023-00274348), all funded by the Korean government. This work was supported by the BK21 FOUR program of the Education and Research Program for Future ICT Pioneers in 2026, through Seoul National University. We also acknowledge an administrative support from SOFT foundry institute.

## Author Contributions

All authors contributed equally to conceiving the idea, developing the code, discussing the results, and preparing the final manuscript.

## Competing Interests

The authors have no conflicts of interest to declare.

## Additional information

**Correspondence and requests for materials** should be addressed to S.Y., X.P., or N.P.

## Figure Legends

**FIG. 1. Magnetic graph model for single-atom cavity QED. a,b,** Schematics for the generalized QRM (**a**) and the eigenstates of an arbitrary atom (**b**). Black arrows in **a** indicate cavity modulations. **c,** Graph model of Eq. (4), with linearly varying self-loops indicated by the graded colour of the nodes (circles). **d,** The FR graph obtained via a local gauge transformation of the graph in **c**.

**FIG. 2. Graph connectivity for different values of $\eta$ at $t = 0$.** Edges with weights greater than 50 % of the global maximum are shown for visibility, and the strength of each weight is indicated by transparency. To suppress errors arising from the finite-dimensional truncation of the Fock space, we employ a large cutoff for the maximum number state: $|n = N_{\max} = 200\rangle$.

**FIG. 3. Stronger coupling indicates a higher cost for separating a large nonmagnetic subgraph. a,** Schematic of the generalized Cheeger's inequality. Sets of green, purple, and red circles denote subgraphs. Scissors mark cuts—setting the edge weight to zero—used to separate subgraphs. Directional dashed arrows and double-headed solid arrows indicate hopping with nonzero and zero phase evolution, respectively. Black and orange arrows denote strong and weak hopping strengths, respectively. **b,** $\lambda_1$ as a function of $\eta$. The red and black dashed lines denote the weak-coupling asymptote and the upper bound on $\lambda_1$, respectively.

**FIG. 4. Phase frustration governs the USC. a,b,** Schematics of subgraphs (coloured nodes) illustrating closed loops for $q = 2$ (**a**) and $q = 3$ (**b**). Black and blue arrows indicate example edges, while the blue ones must be cut to separate the subgraphs. Red curved arrows denote closed loops. **c,d,** Graph measures: $\langle C_q \rangle$ (**c**) and $|W_q|^{1/(2q)}$ (**d**), as functions of $\eta$, for different $q$ values. Both measures are computed over $2\times10^4$ random subgraphs, while $|W_q|^{1/(2q)}$ is evaluated separately for nontrivial (dots) and trivial (dashed lines) phases. **e,f,** Eigenstate analysis: IPR (**e**) and $E_n$ (**f**) as functions of $\eta$. The first 30 eigenstates are plotted.

## References

1. Di Stefano, O., Settineri, A., Macrì, V., Garziano, L., Stassi, R., Savasta, S. & Nori, F.

Resolution of gauge ambiguities in ultrastrong-coupling cavity quantum electrodynamics. *Nat. Phys.* **15**, 803-808 (2019).

2. Ashida, Y., İmamoğlu, A. & Demler, E. Cavity quantum electrodynamics at arbitrary light-matter coupling strengths. *Phys. Rev. Lett.* **126**, 153603 (2021).
3. Akbari, K., Nori, F. & Hughes, S. Floquet engineering the quantum Rabi model in the ultrastrong coupling regime. *Phys. Rev. Lett.* **134**, 063602 (2025).
4. Lednev, M., García-Vidal, F. J. & Feist, J. Lindblad master equation capable of describing hybrid quantum systems in the ultrastrong coupling regime. *Phys. Rev. Lett.* **132**, 106902 (2024).
5. Savasta, S., Di Stefano, O., Settineri, A., Zueco, D., Hughes, S. & Nori, F. Gauge principle and gauge invariance in two-level systems. *Phys. Rev. A* **103**, 053703 (2021).
6. Stokes, A. & Nazir, A. Gauge non-invariance due to material truncation in ultrastrong-coupling quantum electrodynamics. *Nat. Phys.* **20**, 376-378 (2024).
7. Stefano, O. D., Settineri, A., Macrì, V., Garziano, L., Stassi, R., Savasta, S. & Nori, F. Reply to: Gauge non-invariance due to material truncation in ultrastrong-coupling quantum electrodynamics. *Nat. Phys.* **20**, 379-380 (2024).
8. Frisk Kockum, A., Miranowicz, A., De Liberato, S., Savasta, S. & Nori, F. Ultrastrong coupling between light and matter. *Nat. Rev. Phys.* **1**, 19-40 (2019).
9. Forn-Díaz, P., Lamata, L., Rico, E., Kono, J. & Solano, E. Ultrastrong coupling regimes of light-matter interaction. *Rev. Mod. Phys.* **91**, 025005 (2019).
10. Ashhab, S. & Nori, F. Qubit-oscillator systems in the ultrastrong-coupling regime and their potential for preparing nonclassical states. *Phys. Rev. A* **81**, 042311 (2010).
11. Stassi, R., Macrì, V., Kockum, A. F., Di Stefano, O., Miranowicz, A., Savasta, S. & Nori,

F. Quantum nonlinear optics without photons. *Phys. Rev. A* **96**, 023818 (2017).

12. Beaudoin, F., Gambetta, J. M. & Blais, A. Dissipation and ultrastrong coupling in circuit QED. *Phys. Rev. A* **84**, 043832 (2011).
13. Rossatto, D. Z., Villas-Bôas, C. J., Sanz, M. & Solano, E. Spectral classification of coupling regimes in the quantum Rabi model. *Phys. Rev. A* **96**, 013849 (2017).
14. Garziano, L., Settineri, A., Di Stefano, O., Savasta, S. & Nori, F. Gauge invariance of the Dicke and Hopfield models. *Phys. Rev. A* **102**, 023718 (2020).
15. Settineri, A., Di Stefano, O., Zueco, D., Hughes, S., Savasta, S. & Nori, F. Gauge freedom, quantum measurements, and time-dependent interactions in cavity QED. *Phys. Rev. Res.* **3**, 023079 (2021).
16. Mercurio, A., Macrì, V., Gustin, C., Hughes, S., Savasta, S. & Nori, F. Regimes of cavity QED under incoherent excitation: From weak to deep strong coupling. *Phys. Rev. Res.* **4**, 023048 (2022).
17. Watts, D. J. & Strogatz, S. H. Collective dynamics of 'small-world'networks. *Nature* **393**, 440-442 (1998).
18. Wang, D.-W., Cai, H., Liu, R.-B. & Scully, M. O. Mesoscopic superposition states generated by synthetic spin-orbit interaction in Fock-state lattices. *Phys. Rev. Lett.* **116**, 220502 (2016).
19. Wang, Y., Wu, Y.-K., Jiang, Y., Cai, M.-L., Li, B.-W., Mei, Q.-X., Qi, B.-X., Zhou, Z.-C. & Duan, L.-M. Realizing synthetic dimensions and artificial magnetic flux in a trapped-ion quantum simulator. *Phys. Rev. Lett.* **132**, 130601 (2024).
20. Saugmann, P. & Larson, J. Fock-state-lattice approach to quantum optics. *Phys. Rev. A* **108**, 033721 (2023).

21. Yuan, J., Cai, H. & Wang, D.-W. Quantum simulation in Fock-state lattices. *Adv. Phys.: X* **9**, 2325611 (2024).

22. Zhang, X., He, Y., Brugnone, N., Perlmutter, M. & Hirn, M. Magnet: A neural network for directed graphs. *Advances in neural information processing systems* **34**, 27003-27015 (2021).

23. Guo, K. & Mohar, B. Hermitian adjacency matrix of digraphs and mixed graphs. *Journal of Graph Theory* **85**, 217-248 (2017).

24. Pankrashkin, K. Spectra of Schrödinger operators on equilateral quantum graphs. *Letters in Mathematical Physics* **77**, 139-154 (2006).

25. Lange, C., Liu, S., Peyerimhoff, N. & Post, O. Frustration index and Cheeger inequalities for discrete and continuous magnetic Laplacians. *Calculus of variations and partial differential equations* **54**, 4165-4196 (2015).

26. Jerrum, M. & Sinclair, A. *Conductance and the rapid mixing property for Markov chains: the approximation of permanent resolved*. In *Proceedings of the twentieth annual ACM symposium on Theory of computing* 235-244 (1988).

27. Nielsen, M. A. & Chuang, I. *Quantum computation and quantum information* (Cambridge University Press, 2002).

28. Murray, R. M., Li, Z. & Sastry, S. S. *A mathematical introduction to robotic manipulation* (CRC press, 2017).

29. Scully, M. O. & Zubairy, M. S. *Quantum optics* (Cambridge university press, 1997).

30. Emin, D. & Hart, C. Existence of wannier-stark localization. *Phys. Rev. B* **36**, 7353 (1987).

31. Hartmann, T., Keck, F., Korsch, H. & Mossmann, S. Dynamics of Bloch oscillations. *New Journal of Physics* **6**, 2 (2004).

# Supplementary Information for "Magnetic graphs for cavity quantum electrodynamics"

Sunkyu Yu[1†], Xianji Piao[2§], and Namkyoo Park[3*]

[1]Intelligent Wave Systems Laboratory, Department of Electrical and Computer Engineering, Seoul National University, Seoul 08826, Korea

[2]Wave Engineering Laboratory, School of Electrical and Computer Engineering, University of Seoul, Seoul 02504, Korea

[3]Photonic Systems Laboratory, Department of Electrical and Computer Engineering, Seoul National University, Seoul 08826, Korea

E-mail address for correspondence: [†]sunkyu.yu@snu.ac.kr, [§]piao@uos.ac.kr, [*]nkpark@snu.ac.kr

## Supplementary Note S1. Loop formation in the FP graph

## Supplementary Note S1. Loop formation in the FP graph

Figure S1a illustrates loop formation in the FP graph, where $p_{\text{Connect}}$ denotes the probability of forming connected loops across $2 \times 10^4$ random subgraph realizations. When a loop is formed, it can exhibit either a trivial or nontrivial hopping phase. The probabilities of these two cases over all realizations are denoted as $p_{\text{Nontrivial}}$ and $p_{\text{trivial}}$, (Fig. S1b), satisfying $p_{\text{Connect}} = p_{\text{Nontrivial}} + p_{\text{trivial}}$.

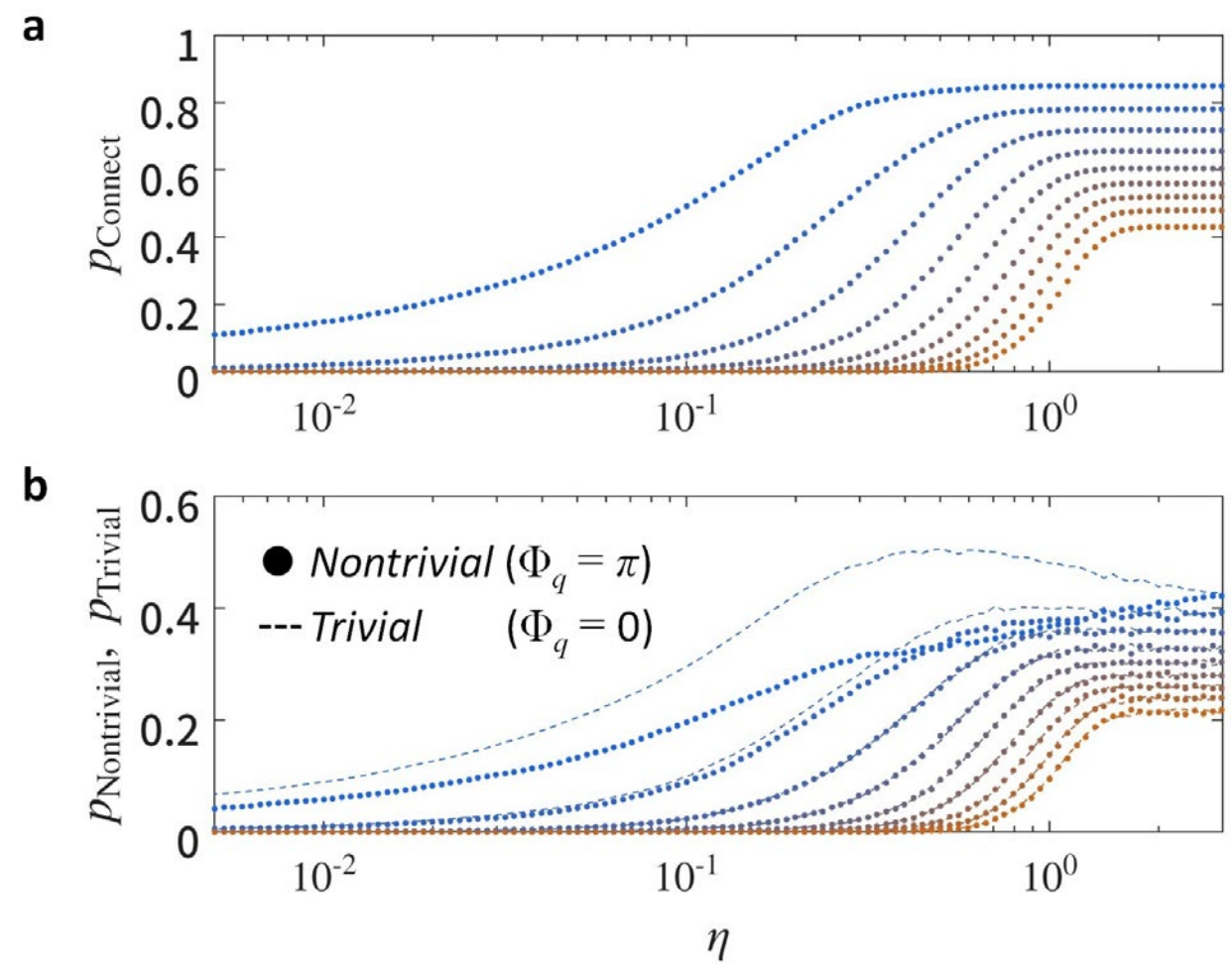

**Supplementary Figure S1. Probabilities of loop formation.** Loop formation probabilities as functions of $\eta$: $p_{\text{Connect}}$ (**a**), and $p_{\text{Nontrivial}}$ and $p_{\text{Trivial}}$ (**b**). All the other parameters are the same as those in Fig. 4 in the main text.